      [1999/12/01 v1.4c Il Nuovo Cimento]
\documentclass{cimento}

\usepackage{graphicx}  
\title{NNLO QCD results for diphoton production at the LHC and the Tevatron\thanks{Contribution to the proceedings
    of the XXVII Rencontres de Physique de la Vall\'ee d'Aoste.}}
\author{L.~Cieri\from{ins:x}\thanks{cieri@fi.infn.it}}
\instlist{\inst{ins:x} INFN, Sezione di Firenze, Via G.Sansone 1, I-50019 Sesto Fiorentino, Florence, Italy.
  }
\PACSes{\PACSit{}{12.38.Bx,  12.38.Cy, 14.80.Bn}}
\begin{document}

\maketitle

\begin{abstract}
We consider direct diphoton
production in hadron collisions. We compute the next-to-next-to-leading
order (NNLO) QCD radiative corrections at the fully-differential level.
Our calculation is based on the $q_T$ subtraction formalism and it is implemented
in a parton level Monte Carlo program, which allows the user to apply arbitrary kinematical cuts on the
final-state photons and the associated jet activity, and to compute the
corresponding distributions in the form of bin histograms. We present selected
numerical results related to Higgs boson searches and diphoton studies performed at the LHC and the Tevatron, and we show how
the NNLO corrections to diphoton production are relevant to understand the main background
of the decay channel $H\rightarrow\gamma \gamma$ of the Higgs boson $H$.
\end{abstract}

\section{Introduction}
\noindent
The measurements of the production cross section of two energetic isolated photons (diphotons) in high energy hadron collisions is important for testing Standard Model (SM) predictions in the domain of searches for undiscovered particles and new physics. Understanding the reaction mechanisms in complicated environment formed in such collision is a challenge for perturbative Quantum Chromodinamics (pQCD) calculations. Photons originating from collisions of hadrons (``direct'' or ``prompt'' photons) are an ideal probe for testing these calculations because they do not interact with other final state particles, and their energies and directions can be measured with high precision in modern electromagnetic calorimeters. 

Hadronic production of isolated diphotons is not only relevant to QCD studies. 
The origin of the Electroweak symmetry breaking is currently
being
investigated at the LHC by searching for the Higgs boson and
 studying its properties. 
Recent results in the search for the SM Higgs Boson at the LHC indicates the observation of a new particle~\cite{cha:2012gu,aad:2012gk}, which is a neutral boson with a mass $M\sim 125$~GeV. 
In this spectacular new observation, as well as in previous searches and studies, the preferred search mode involves Higgs boson production via gluon fusion followed by the rare decay into a pair of photons. 
Therefore, it is essential to count on an accurate theoretical description of
the various kinematical distributions associated to the production of pairs of
prompt photons with large invariant mass. An improved knowledge of the SM background will help the development of more powerful search strategies and studies for this particle. Moreover, diphoton production is also an irreducible background for some new physics searches, beyond the standard model. 

All precedent tasks require a detailed study of the various kinematical distributions associated to the production of pairs of prompt photons with large invariant mass, and therefore their comparison with an accurate theoretical description. Such accurate comparison requires detailed computations of radiative corrections, which implies finding methods and techniques to practically
achieve the cancellation of infrared (IR) divergences that appear at 
intermediate steps of the calculations.


\section{Theoretical content and outline of the computation}
In this paper we are interested in the process $pp\rightarrow \gamma \gamma X$ (and the related process $p\bar{p}\rightarrow \gamma \gamma X$) which at the lowest order occurs \textit{via} the quark annihilation subprocess $q\bar{q}\rightarrow \gamma\gamma$. The QCD corrections at 
the NLO in the strong coupling $\alpha_S$ involve the  
quark annihilation channel and a new partonic channel, \textit{via} the
subprocess $qg \rightarrow \gamma \gamma q$. These corrections have been
computed and implemented in the fully-differential Monte Carlo codes
\texttt{DIPHOX} \cite{Binoth:1999qq}, \texttt{2gammaMC} \cite{Bern:2002jx} and 
\texttt{MCFM} \cite{Campbell:2011bn}. A calculation that includes the effects of
 transverse-momentum resummation is implemented in 
\texttt{RESBOS} 
\cite{Balazs:2007hr}.

At the NNLO, 
the $gg$ channel starts to contribute,
and
the large gluon--gluon luminosity makes this channel potentially sizeable. 
Part of the contribution from this channel,
the so called {\it box contribution} 
, was computed long ago \cite{Dicus:1987fk} and 
its size turns out to be comparable  
to the lowest-order result; for this reason,
the {\it box contribution} is customarily included in all the NLO computations of diphoton production.
The next-order gluonic corrections to the {\it box contribution}
(which are part of the N$^3$LO QCD corrections to diphoton production)
were computed in ref.~\cite{Bern:2002jx} and found 
to have a moderate quantitative effect on the result of the `NLO+box' calculation.

The  \texttt{2$\gamma$NNLO} code incorporates, for first time, the complete description for direct diphoton production at second order (NNLO) in the strong coupling constant $\alpha_S$. In this sense, it is the first fully consistent inclusion of the box contribution, since we are considering all the partonic amplitudes at $\mathcal{O}(\alpha_S^2)$ for the $gg$ channel for the first time. 
At the second order in $\alpha_S$, we also have the first correction to the $qg$ channel~\cite{Barger:1989yd,Bern:1994fz}, which is the dominant channel for this process~\cite{Catani:2011qz} (at NLO and at NNLO), 
Finally we incorporate the second order corrections to the Born subprocess ($q\bar{q}\rightarrow\gamma\gamma$), which requires the knowledge of {\it double real radiation} contributions~\cite{Barger:1989yd}, real-virtual corrections~\cite{Bern:1994fz} to the Born subprocess and the {\it two-loops} correction terms~\cite{ Anastasiou:2002zn} to the LO subprocess.


Besides the {\it direct} photon production from the hard subprocess, photons can also
arise from fragmentation subprocesses of QCD partons. The computation of
fragmentation subprocesses requires (poorly known)
non-perturbative information, in the form of 
parton  
fragmentation functions of the photon.
The complete NLO single- and double-fragmentation contributions are implemented in \texttt{DIPHOX}
\cite{Binoth:1999qq}. 

The effect of the fragmentation contributions 
is sizeably reduced by the photon isolation criteria that are 
necessarily
applied in hadron collider experiments to suppress the very large irreducible
background (\textit{e.g.}, photons that are faked by jets or produced by hadron decays). 
The standard cone isolation and the `smooth' cone isolation proposed
by Frixione \cite{Frixione:1998jh} are two of these criteria. The standard cone
isolation is easily implemented in experiments, but it only suppresses a fraction of the 
fragmentation contribution.
The smooth cone isolation (formally) eliminates the entire fragmentation 
contribution, but its experimental implementation (at least in its original form) is complicated~\footnote{There is activity in the experimental implementation~\cite{Binoth:2010ra,ATLASBLAIROP,wielers} of the discretized version of the Frixione isolation criterion. An experimental implementation of the smooth isolation criterion
was done by the OPAL collaboration~\cite{Abbiendi:2003kf}.} by the finite granularity of the LHC and Tevatron detectors~\cite{Binoth:2010ra}.

However it is possible (\textit{i.e.} it has physical meaning) to compare theoretical descriptions obtained using the smooth cone isolation criterion and data taken with the standard criterion,
because always a cross section obtained using the Frixione isolation criterion is always a lower bound for a cross section in which the standard criterion was implemented, if we use the same isolation parameters for both criteria.

\subsection{The transverse momentum subtraction formalism at NNLO}
The implementation of the separate scattering amplitudes in a complete NNLO (numerical) calculation is severely complicated by the presence of infrared (IR) divergences, that occur at intermediated stages of the computation, and for this reason, it is mandatory the use of a prescription to handle an cancel them inside a Monte Carlo code.
We consider the inclusive hard-scattering reaction
\begin{equation}
\label{one}
h_1+h_2\to \gamma\gamma +X \;\;,
\end{equation}
where the collision of the two hadrons, $h_1$ and $h_2$,
produces the diphoton system 
$F \equiv \gamma\gamma$ with high invariant mass $M_{\gamma \gamma}$.
The evaluation of the
NNLO corrections to this process requires the knowledge 
of the corresponding partonic scattering amplitudes
with $X=2$~partons (at the tree level \cite{Barger:1989yd}), $X=1$~parton (up 
to the one-loop level \cite{Bern:1994fz})
and no additional parton (up to the two-loop level \cite{Anastasiou:2002zn})
in the final state.
The implementation of the separate scattering amplitudes in a complete
NNLO (numerical) calculation is severely complicated by 
the presence of infrared (IR) divergences that occur at intermediate stages. 
The $q_T$ subtraction formalism \cite{Catani:2007vq} is a method that handles
and cancels these unphysical IR divergences up to the NNLO.
The formalism applies to generic hadron collision processes that involve
hard-scattering production of a colourless high-mass system $F$.
 Within that framework~\cite{Catani:2007vq}, the corresponding cross section is written as:
\begin{equation}
\label{main}
d{\sigma}^{F}_{(N)NLO}={\cal H}^{F}_{(N)NLO}\otimes d{\sigma}^{F}_{LO}
+ \big[ d{\sigma}^{F+{\rm jets}}_{(N)LO}
- d{\sigma}^{CT}_{(N)LO}\big]\;\;,
\end{equation}
where $d{\sigma}^{F+{\rm jets}}_{(N)LO}$ represents the cross section for the
production of the system $F$ plus jets at (N)LO accuracy~\footnote{In the case of
diphoton production, the NLO calculation of 
$d{\sigma}^{\gamma\gamma+{\rm jets}}_{NLO}$ was performed in 
ref.\cite{DelDuca:2003uz}.}, and
$d{\sigma}^{CT}_{(N)LO}$ is a (IR subtraction) counterterm whose explicit expression \cite{Bozzi:2005wk}
is obtained from the resummation program of the logarithmically-enhanced
contributions to $q_T$ distributions. 
The `coefficient' ${\cal H}^{F}_{(N)NLO}$, which also compensates for the subtraction
of $d{\sigma}^{CT}_{(N)LO}$,
corresponds to the (N)NLO truncation of the process-dependent perturbative function
\begin{equation}
{\cal H}^{F}=1+\frac{\alpha_{\mathrm{S}}}{\pi}\,
{\cal H}^{F(1)}+\left(\frac{\alpha_{\mathrm{S}}}{\pi}\right)^2
{\cal H}^{F(2)}+ \dots \;\;.
\end{equation}
The NLO calculation  of $d{\sigma}^{F}$ 
requires the knowledge
of ${\cal H}^{F(1)}$, and the NNLO calculation also requires ${\cal H}^{F(2)}$. The general 
structure of ${\cal H}^{F(1)}$
is explicitly known~\cite{deFlorian:2000pr}; exploiting the explicit results of ${\cal H}^{F(2)}$ for Higgs
\cite{Catani:2007vq,Catani:2011kr} and vector boson \cite{Catani:2009sm,Catani:2012qa} 
production, we have generalized the process-independent relation of ref.~\cite{deFlorian:2000pr} to 
the calculation of the NNLO coefficient 
${\cal H}^{F(2)}$.

\section{The phenomenology of direct diphoton production}


\subsection{The isolation criteria}
\label{Sect:Frix}

Both sides, the experimental and the theoretical one suffer distinctive types of difficulties when they apply an isolation criterion. 

\noindent
Prompt photons can be produced according to two possible mechanisms, one of them being
a fragmentation mechanism. 

The isolation criterion used by collider experiments is schematically as follows. A photon is said to be isolated if, in a cone of radius $R$
in rapidity and 
azimuthal angle around the photon direction, the amount of deposited hadronic 
transverse energy $\sum E_{T}^{had}$ is smaller than some value 
$E_{T \, max}$ 
chosen by the experiment:  
\begin{equation}\label{standardcriterion}     
\sum E_{T}^{had} \leq E_{T \, max} \;\;\;\; \mbox{inside} \;\;\;\;      
\left( y - y_{\gamma} \right)^{2} +    
\left(  \phi - \phi_{\gamma} \right)^{2}  \leq R^{2}  \;,    
\end{equation} 
where $E_{T \, max}$ can be either, a fixed value
or a fraction of the transverse momentum of the photon ($p_T^{\gamma}\epsilon$, where typically $0 < \epsilon \leq 1$). This is the so-called standard cone isolation criterion. In addition to the rejection of the background of secondary  photons, this 
isolation cuts also affect the prompt-diphoton cross section itself, 
in particular by reducing the effect of fragmentation.
However a tight isolation cut also has the undesirable effect of making the theoretical prediction unstable~\cite{Catani:2002ny}, due to the restriction of the available phase-space for parton emission.


There exist an alternative to the standard criterion: the criterion proposed by Frixione in ref.~\cite{Frixione:1998jh}
(see also refs.~\cite{Frixione:2000gr,Catani:2000jh}). This criterion modifies eq.~(\ref{standardcriterion}) in the following way
\begin{equation}\label{frixcriterion}     
\sum E_{T}^{had} \leq E_{T \, max}~\chi(r)\;, \;\;\;\; \mbox{inside any} \;\;\;\;      
r^{2}=\left( y - y_{\gamma} \right)^{2} +    
\left(  \phi - \phi_{\gamma} \right)^{2}  \leq R^{2}  \;.    
\end{equation}  
with a suitable choice for the function $\chi(r)$. This function has to vanish smoothly when its argument goes to zero ($\chi(r) \rightarrow 0 \;,\; \mbox{if} \;\; r \rightarrow 0\,$), and it has to verify \mbox{$\; 0<\chi(r)< 1$}, if \mbox{$0<r<R\,\,.$} One possible choice is,
\begin{equation}
\label{Eq:chinormal}
\chi(r) = \left( \frac{1-\cos (r)}{1-\cos R} \right)^{n}\;,
\end{equation}
where $n$ is typically chosen as $n=1$. This means that, closer to the photon, less hadronic activity is permitted inside the cone. At $r=0$, when the parton and the photon are exactly collinear, the energy deposited inside the cone is required to be exactly equal to zero, and the fragmentation component (which is a purely collinear phenomenon in perturbative QCD) vanishes completely. Since no region of the phase space is forbidden, the cancellation of soft gluon effects takes place as in ordinary infrared-safe cross sections. This is the advantage of this criterion: it kills all the fragmentation
component in an infrared-safe way.
We can also say, comparing eqs.~\ref{standardcriterion} and \ref{frixcriterion}, that both criteria coincides at the outer cone ($r=R$, $\chi(R)=1$), and due to the presence of the $\chi(r)$ function which verifies $0\leq\chi(r)\leq 1$, the smooth cone isolation criterion is more restrictive than the standard one; and for this reason, we expect smaller cross sections, when we use the Frixione criterion than when we implement the standard one, both theoretically and experimentally, if the same parameters~\footnote{\textit{I.e}, the same ($E_{T \, max},R$) or ($\epsilon,R$).} are used in both criteria,
\begin{equation}
\sigma_{Frix}\{R,E_{T~max}\}\leq \sigma_{Stand}\{R,E_{T~max}\}\,\,.
\end{equation}
The smooth behaviour of the $\chi(r)$ function is the main obstacle to implement the Frixione isolation criterion into the experimental situation. First, because of the finite size of the calorimeter cells used to measure the electromagnetic shower, the smooth cone criterion must be applied only beyond a minimum distance of approximately $0.1$ (in $\{ \Delta \eta, \Delta \phi \}$ plane). This allows a contribution from fragmentation in the innermost cone and we have to check to which extent the fragmentation component is still suppressed. In addition, the transverse energy in the experimental isolation cone is deposited in discrete cells of finite size. 
The continuity criterion, initially proposed by Frixione has thus been replaced by a discretized version (concerning the experimental side) consisting of a finite number of nested cones, together with the collection of corresponding maximal values for the transverse energy allowed inside each of these cones.
\subsection{Quantitative results}
We have performed our fully-differential NNLO calculation~\cite{Catani:2011qz} of diphoton production
according to eq.~(\ref{main}).
The NNLO computation is encoded
in a parton level
Monte Carlo program \texttt{2$\gamma$NNLO}, in which
we can implement arbitrary IR safe cuts on the final-state
photons and the associated jet activity. 
We concentrate on the direct production of diphotons, and 
we rely on the smooth cone isolation criterion~\cite{Frixione:1998jh}.
We use the MSTW 2008~\cite{Martin:2009iq} sets of parton distributions, with densities and $\alpha_{\mathrm{S}}$ evaluated at each corresponding order,
and we consider $N_f=5$ massless quarks/antiquarks and gluons in 
the initial state. The default
renormalization ($\mu_R$) and factorization ($\mu_F$) scales are set to the value
of the invariant mass of the diphoton system,
$\mu_R=\mu_F = M_{\gamma\gamma}$. The QED coupling constant $\alpha$ is fixed to $\alpha=1/137$. To present some quantitative results, we consider diphoton production
at the LHC ($\sqrt{s}=14$~TeV). We apply typical kinematical cuts used by ATLAS and CMS 
Collaborations in their Higgs boson search studies. We
require the harder and the softer photon to have transverse momenta $p_T^{\rm harder}\geq
40$~GeV and $p_T^{\rm softer}\geq 25$~GeV, respectively.
The rapidity of both photons is restricted to $|y_\gamma| \leq 2.5$, and
the invariant mass of the 
diphoton system is constrained to 
lie in the range $20 \,{\rm GeV}\leq M_{\gamma\gamma} \leq 250\,{\rm GeV}$.
The isolation parameters
are set to the values $\epsilon_\gamma=0.5$, $n=1$ and $R=0.4$. 
Performing the QCD computation, we observe \cite{Catani:2011qz} that 
the value of the cross section remarkably increases with the perturbative order
of the calculation. This increase is mostly due to the use of {\it very
asymmetric} (unbalanced) cuts on the photon transverse momenta. At the LO,
kinematics implies that the two photons are produced with equal transverse
momentum and, thus, both photons should have $p_T^{\gamma}\geq 40$~GeV. 
At higher orders, the final-state radiation of additional partons opens a new
region of the phase space, 
where $40$~GeV $\geq p_T^{\rm softer}\geq 25$~GeV. Since photons can copiously be
produced with small transverse momentum~\cite{Catani:2011qz}, the cross section receives a sizeable
contribution from the enlarged phase space region. This effect is further
enhanced by the opening of a new large-luminosity partonic channel at each
subsequent perturbative 
order. 
\begin{figure}
\includegraphics[width=0.48\textwidth,height=0.42\textwidth]{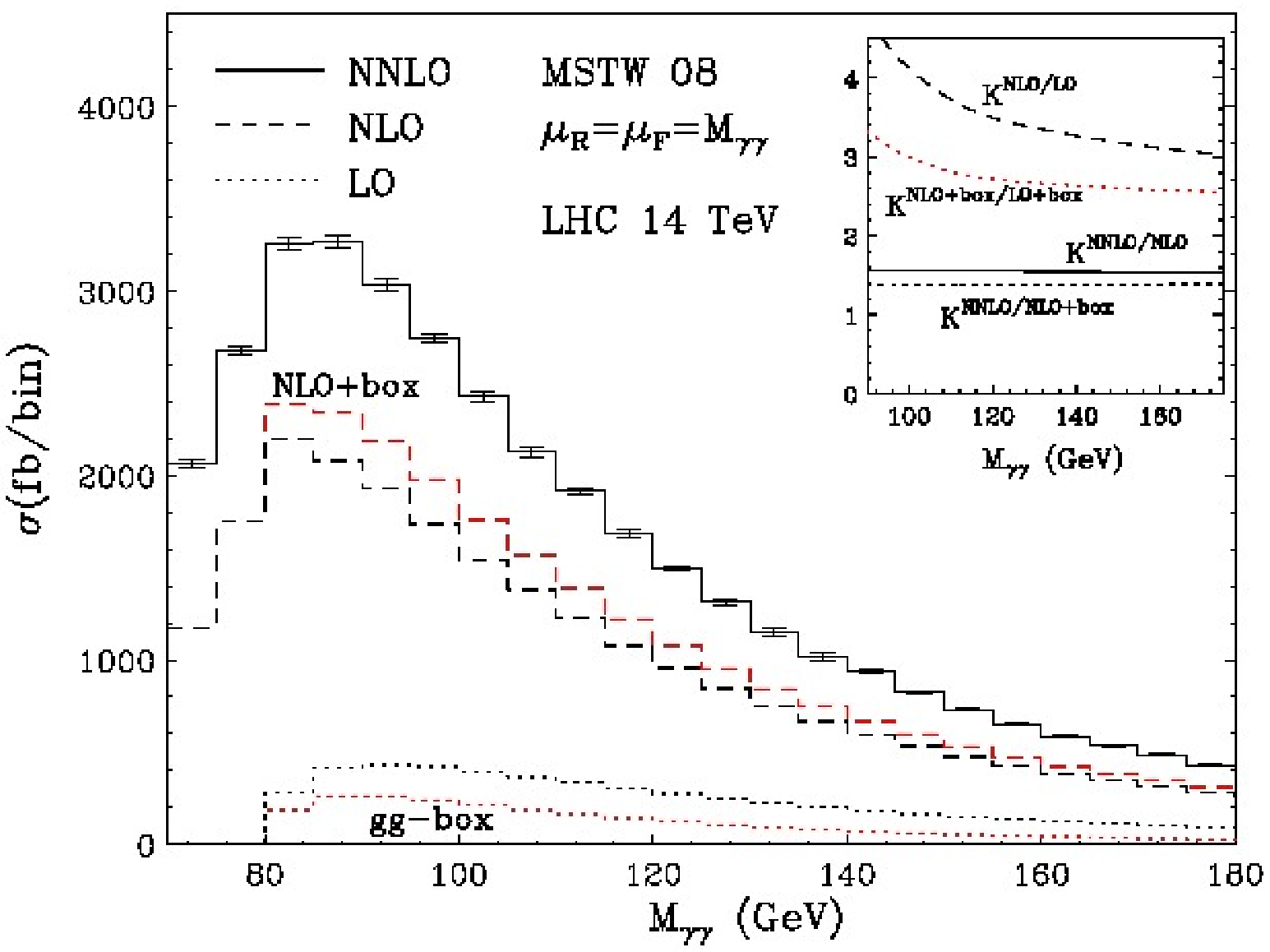}\hfill
\includegraphics[width=0.48\textwidth,height=0.42\textwidth]{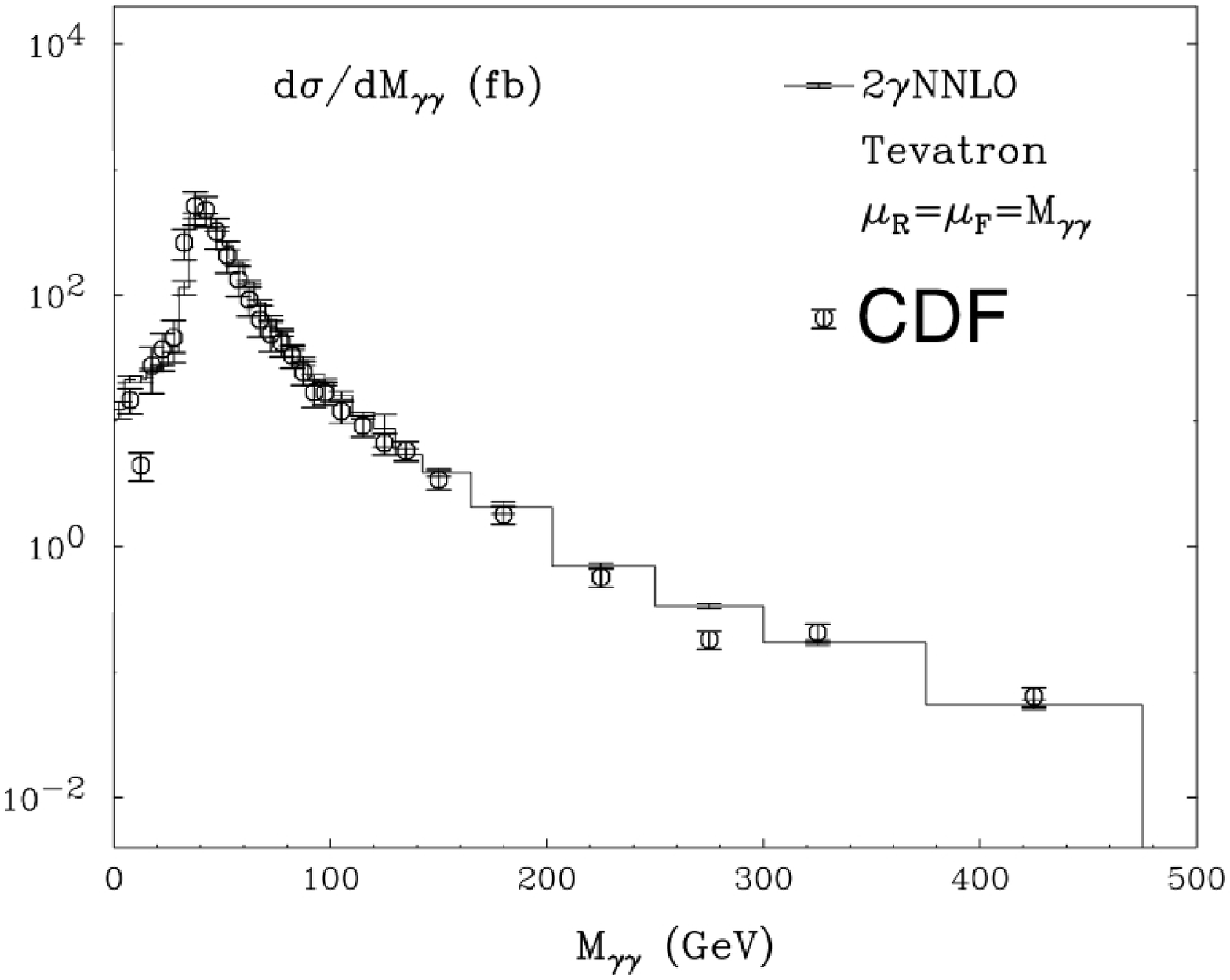}
\caption{Left: Invariant mass distribution of the photon pair
at the LHC 
($\sqrt{s}=14$~{\rm TeV}): LO (dots), NLO (dashes) and NNLO (solid) results. We
also present the results of the box and NLO+box contributions. The inset plot
shows the 
corresponding {\rm K}-factors. Right: Diphoton cross section as a function of the invariant mass of the two photons. Data from CDF~\cite{Aaltonen:2011vk} ($\sqrt{s}=1.96$~TeV) are compared to the NNLO calculation.}
\label{fig:mass}
\end{figure}
In fig.~\ref{fig:mass} (left) we show the LO, NLO and NNLO invariant mass
distributions at the default scales. 
We also plot the gluonic {\it box
contribution} (computed with NNLO parton distributions) and its sum with the full NLO result.
 The inset plot shows the K-factors defined as 
the ratio of the cross sections at two subsequent perturbative orders.
We note that ${\rm K}^{NNLO/NLO}$ is sensibly smaller than ${\rm K}^{NLO/LO}$,
and this fact indicates 
an improvement in the convergence of the perturbative expansion.
We find that about 30\% of the NNLO corrections is due to the $gg$ channel (the 
{\it box contribution} is responsible for more than half of it), while almost
60\% still arises from the next-order corrections to the $qg$ channel.
The NNLO calculation includes the perturbative corrections from the entire phase
space region 
(in particular, the next-order correction to the dominant $qg$ channel)
and the contributions from all possible partonic channels (in particular,
a fully-consistent treatment of the {\it box contribution} to 
the $gg$ channel~\footnote{The calculation \cite{Bern:2002jx} of the next-order
gluonic corrections to the  {\it box contribution} indicates an increase of 
the NNLO result by less than 10\% if $M_{\gamma\gamma} \geq 100$~GeV.}).
Owing to these reasons, the NNLO result can be considered a reliable estimate of
direct diphoton production, although further studies (including independent
variations of $\mu_R$ and $\mu_F$, and detailed analyses of kinematical distributions)
are necessary to quantify the NNLO theoretical uncertainty.
In figure~\ref{fig:mass} (right), we present the invariant mass distribution for 
diphoton production at the Tevatron ($\sqrt{s}=1.96$~TeV) calculated at NNLO, compared with a measurement performed by CDF~\cite{Aaltonen:2011vk}. The acceptance criteria in this case require: $p_T^{\rm harder}\geq
17$~GeV and $p_T^{\rm softer}\geq 15$~GeV. The rapidity of both photons is restricted to \mbox{$|y_\gamma| \leq 1$}. The isolation parameters are set to the values $E_{T~max}=2$~GeV, $n=1$ and $R=0.4$, and the minimum angular separation between the two photons is $R_{\gamma\gamma}=0.4$.
Though in this case the increase from the LO to the NLO result is considerably smaller than at the LHC~\cite{Catani:2011qz}, the NNLO QCD corrections still improve remarkably the theoretical description of the CDF data, in particular in the low mass region ($M_{\gamma\gamma} \leq 2p_T^{harder}=34$~GeV).
\begin{figure}
\includegraphics[width=0.49\textwidth,height=0.55\textwidth]{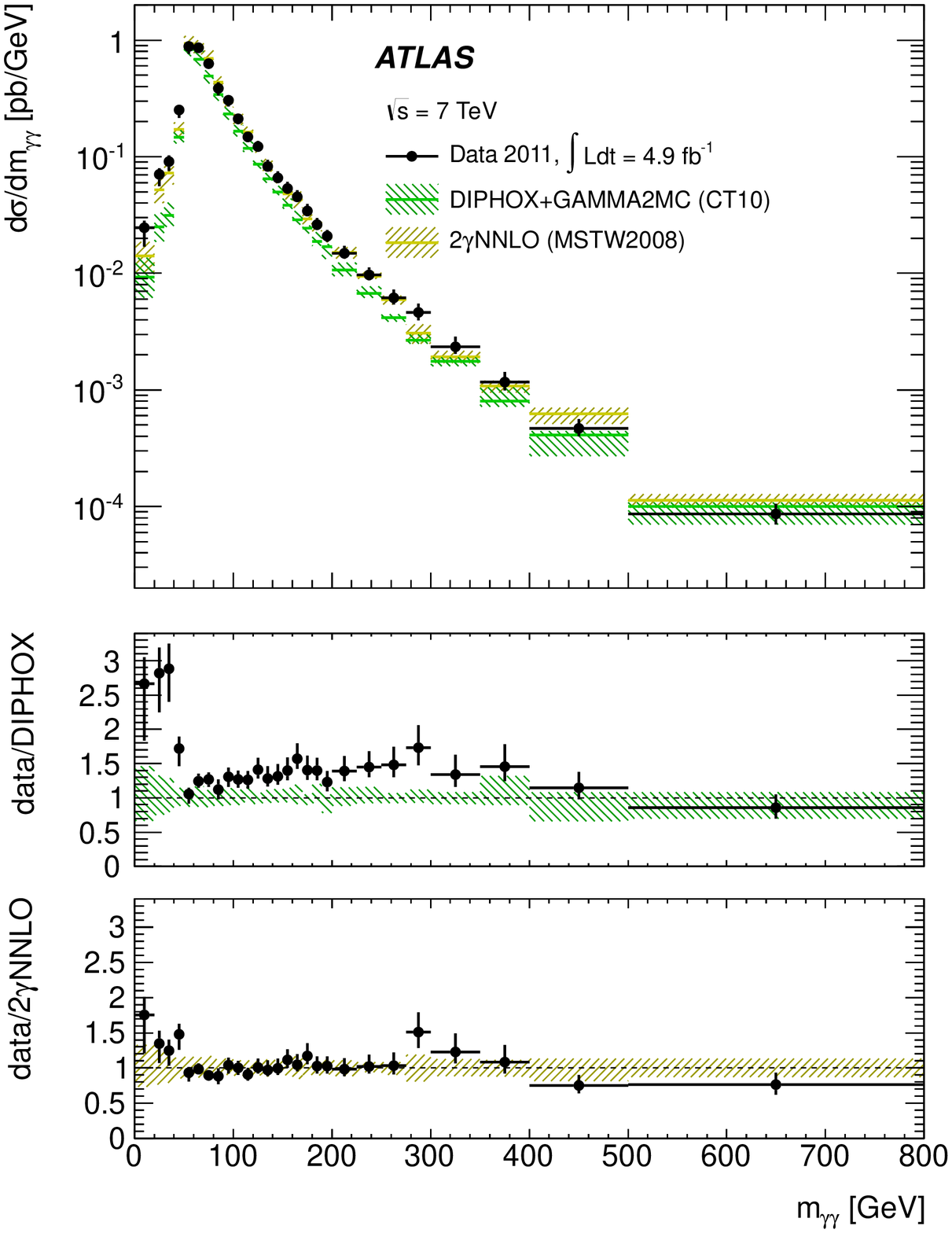}\hfill
\includegraphics[width=0.49\textwidth,height=0.55\textwidth]{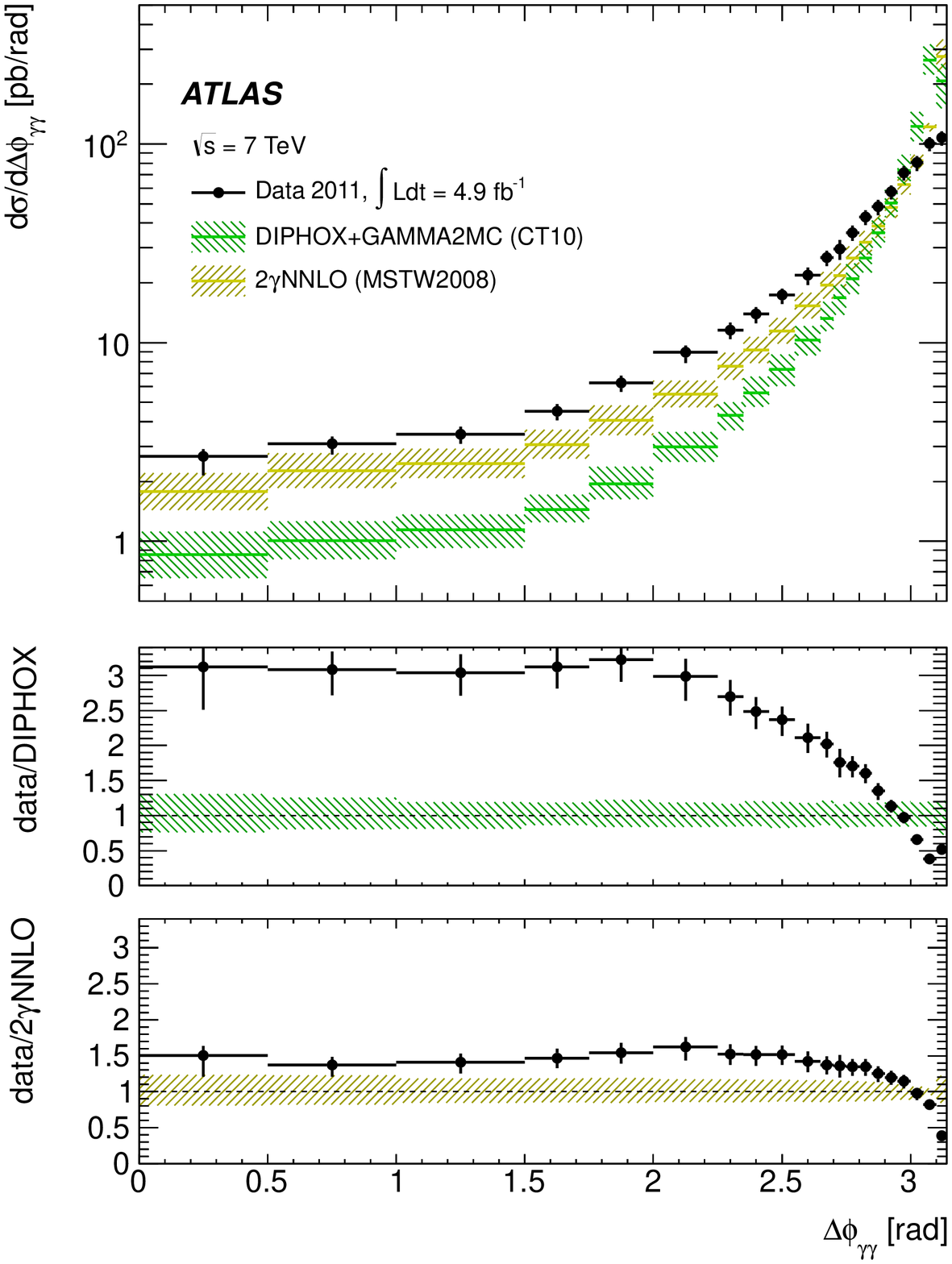}
\caption{ Diphoton 
cross section as a function of the invariant mass $M_{\gamma\gamma}$ (left) and the azimuthal separation of
  the two photons (right). Data from ATLAS~\cite{Aad:2012tba} ($\sqrt{s}=7$~{\rm TeV}) are
  compared to the NNLO calculation~\cite{Catani:2011qz}.}
\label{fig:ATLASmgg}
\end{figure}
Recent results from the LHC~\cite{Chatrchyan:2011qt,Aad:2011mh} and the
Tevatron \cite{Aaltonen:2011vk} show some discrepancies between the data
and NLO theoretical calculations of diphoton production. Basically, discrepancies were found in
kinematical regions where the NLO calculation is {\em effectively} a LO theoretical 
description of the process. Such 
phase space regions (away from the back-to-back configuration) are accessible at 
NLO for the first time, due to the final-state radiation of the 
additional parton~\footnote{The low-mass
region ($M_{\gamma\gamma}\leq 80GeV$) in figure \ref{fig:mass} also belongs to this case.}. 

Figure~\ref{fig:ATLASmgg} 
shows a measurement by ATLAS~\cite{Aad:2012tba}, of the diphoton
cross section as a function of the invariant mass $M_{\gamma\gamma}$ (left) and the azimuthal angle $\Delta \phi_{\gamma\gamma}$ (right) between the
photons. The data are compared with NLO~\cite{Binoth:1999qq} and NNLO calculations~\cite{Catani:2011qz}.
The acceptance criteria used in this analysis ($\sqrt{\rm s}=7$~TeV) require: $p_T^{\rm harder}\geq
25$~GeV and $p_T^{\rm softer}\geq 22$~GeV.
The rapidity of both photons is restricted to $|y_\gamma|<1.37$ and \mbox{$1.52<|y_\gamma| \leq 2.37$}. The isolation parameters
have the values $E_{T~max}=4$~GeV, $R=0.4$ and the minimum angular separation between the two photons is $R_{\gamma\gamma}=0.4$.
The histograms in fig.~\ref{fig:ATLASmgg} show
that the NNLO QCD results remarkably improve the theoretical description of the ATLAS data throughout the 
entire range of  $M_{\gamma\gamma}$ and $\Delta \phi_{\gamma\gamma}$.

We note that the ATLAS and CDF data are obtained by using the standard cone isolation criterion.
Since the smooth isolation criterion used in our calculation (we apply eq.~(\ref{frixcriterion}) for all cones with $r\leq R$) is stronger than the photon
isolation used by ATLAS and CDF, we remark that our NLO and NNLO results cannot overestimate the corresponding
theoretical results for the experimental isolation criterion. 
The results illustrated in this contribution show that the NNLO description of diphoton 
production is essential to understand the 
phenomenology associated to this process, and therefore, the NNLO calculation is a relevant tool 
to describe the main background for Higgs boson searches and studies.
\acknowledgments
I would like to thank Stefano Catani and Daniel de Florian for helpful comments, and to Yamila Rotstein for her support. This work was 
supported by the INFN and the Research Executive Agency (REA) of the European Union under the Grant Agreement number PITN-GA-2010-264564 (LHCPhenoNet).

\end{document}